\documentclass[12pt, reqno]{article}

\usepackage{jheppub}
\usepackage{amssymb}
\usepackage{amsmath}
\usepackage[usenames,dvipsnames]{xcolor}
\usepackage{epsfig}
\usepackage{dcolumn}
\usepackage{tikz}
\usetikzlibrary{shapes.geometric, arrows}
\usepackage{upgreek}
\usepackage{setspace}
\usepackage{enumitem}
\usepackage{array,multirow,bigdelim,arydshln}
\usepackage{appendix}
\usepackage{xparse}
\NewDocumentCommand{\binomial}{omm}
 {%
  \genfrac(){0pt}{}{#2}{#3}%
  \IfValueT{#1}{_{\!#1}}%
 }
\NewDocumentCommand{\eulerian}{omm}
 {%
  \genfrac<>{0pt}{}{#2}{#3}%
  \IfValueT{#1}{_{\!#1}}%
 }

\usepackage{latexsym}
\usepackage{tikz}
\title{Minimal Basis in Four Dimensions and Scalar Blocks}
\author[a]{Freddy Cachazo}
\author[a,b]{and Guojun Zhang}
\affiliation[a]{Perimeter Institute for Theoretical Physics, Waterloo, ON N2L 2Y5, Canada}
\affiliation[b]{Department of Physics $\&$ Astronomy, University of Waterloo, Waterloo, ON N2L 3G1, Canada}
\emailAdd{fcachazo@pitp.ca}
\emailAdd{gzhang2@pitp.ca}
\abstract{We find a construction that expresses any tree-level $n$-particle ${\rm N^{k-2}MHV}$ color-ordered partial amplitude in gauge theory as a linear combination of a basis of dimension $\eulerian{n-3}{k-2}$. Here $\eulerian{p}{q}$ denotes the $(p,q)$ Eulerian number. The coefficients of the expansion are independent of the helicities of the particles. This basis is a four-dimensional refinement of the $(n-3)!$-element BCJ basis which is valid in any number of dimensions. The construction uses a new kind of objects which we call {\it scalar blocks}. Here we initiate the study of these objects. Scalar blocks provide an ``${\rm N^{k-2}MHV}$ sector" decomposition of a bi-adjoint scalar amplitude in four dimensions. As byproducts of the construction, we also find an intrinsically four-dimensional version of KLT relations for gravity amplitudes.}
\begin{document}
\maketitle
\def \tr {\nonumber\\}
\def \la  {\langle}
\def \ra {\rangle}
\def\hset{\texttt{h}}
\def\gset{\texttt{g}}
\def\sset{\texttt{s}}
\def \be {\begin{equation}}
\def \en {\end{equation}}
\def \bes {\begin{eqnarray}}
\def \ens {\end{eqnarray}}
\def \k {\kappa}
\def \h {\hbar}
\def \r {\rho}
\def \l {\lambda}


\numberwithin{equation}{section}

\section{Introduction and Summary of Results}

In 2003 Witten proposed a reformulation of the tree-level S-matrix of ${\cal N}=4$ super-Yang-Mills as an integral that contains the moduli space of a punctured sphere \cite{witten}. This formulation was studied and developed by Roiban, Spradlin and Volovich in 2004 \cite{RSV}. The Witten-RSV formulas were eventually extended to ${\cal N}=8$ supergravity amplitudes in \cite{Cachazo and Geyer, Skinner and Cachazo,Cachazo:2012pz}.

The Witten-RSV formula is an integral that localizes on solutions to a system of polynomial equations. The equations depend on the R-charge sector under consideration. In the ${\rm N^{k-2}MHV}$ sector, the equations for an $n$-particle amplitude have $\eulerian{n-3}{k-2}$ solutions. Here $\eulerian{p}{q}$ is the $(p,q)$ Eulerian number\footnote{The definition of Eulerian numbers is given recursively $\eulerian{p}{q} = (p-q)\eulerian{p-1}{q-1}+(q-1)\eulerian{p-1}{q}$ with $\eulerian{1}{0}=1$ and $\eulerian{p}{q}=0$ for $p\leq q$.}. In 2012, Geyer and the first author found a relation between the solutions and the Kawai-Lewellen-Tye (KLT) relations \cite{KLT} which is known as KLT orthogonality \cite{Cachazo and Geyer}. KLT orthogonality was proven in \cite{CHY} and also led to an algebraic proof of the set of quadratic relations discovered in 2010 by Bjerrum-Bohr, Damgaard, Feng and Sondergaard (BDFS) \cite{BDFS}.

In a seemingly different line of research, Bern, Carrasco and Johannson \cite{Bern:2008qj} found that the basis of $n$-particle color-ordered partial amplitudes in gauge theory can be further reduced from the $(n-2)!$ dimensional Kleiss-Kuijf (KK) \cite{KK} basis down to $(n-3)!$ if one allows linear relations with coefficients that depend on Mandelstam invariants, $s_{ab}=(k_a+k_b)^2$, and are independent of the helicity of the particles in the amplitudes. In fact, the BCJ basis is valid for amplitudes in any number of dimensions where the separation of amplitudes into sectors does not exist. These relations were later derived using string theory methods in \cite{string1,string2,string3}.

A natural question that arises from the BCJ result and the BDFS quadratic relations is the existence of a smaller basis when restricting to four dimensions and to a particular helicity sector \cite{BDFS, He-Feng}. The simplest hint that such a thing is possible is the stunning simplicity of MHV amplitudes. Slightly extending the criteria for allowed coefficients from Mandelstam invariants in the BCJ relations to also include combinations of scalars and pseudo-scalars, one can conclude that the basis of amplitudes in the MHV sector is only one-dimensional! More explicitly, any MHV $n$-particle partial amplitude can be trivially written in terms of the one with the canonical ordering as
\be\label{intro1}
A^{\rm YM}_{n,{\rm MHV}}(\alpha_1,\alpha_2,\ldots, \alpha_n ) = \frac{\langle 1~2\rangle[\alpha_2\alpha_3]\cdots \langle n~1\rangle[\alpha_n\alpha_1] }{s_{\alpha_1\alpha_2}s_{\alpha_2\alpha_3}\cdots s_{\alpha_n\alpha_1}} A^{\rm YM}_{n,{\rm MHV}}(1,2,\ldots ,n),
\en
where the coefficient does not carry helicity and hence it is a combination of a scalar and a pseudo-scalar. Of course, if Parke and Taylor had not provided their beautiful formula \cite{Parke-Taylor}, discovering a relation such as \eqref{intro1} would have been hard and very surprising.

In \cite{BDFS, He-Feng}, NMHV amplitudes of six particles were studied and two relations were found among the six amplitudes in the BCJ basis. These relations came from the quadratic relations using MHV and $\overline{{\rm MHV}}$ as the second set of amplitudes. The first relation is
\be
[2|3+4+5|6\rangle[3|4+5|6\rangle[4|5|6\rangle A_{6,{\rm NMHV}}^{\rm YM}(1,2,3,4,5,6) + {\cal P}(234)  = 0
\en
while the second is obtained by parity-conjugating the coefficients. These two relations can then be used to reduce the number of independent partial amplitudes from $(6-3)!=6$ down to $4$.

The observations made above give a one-dimensional basis for MHV ($k=2$) and for $\overline{{\rm MHV}}$ ($k=n-2$) while a four-dimensional basis for $n=6$ and $k=3$. These are exactly the values of the corresponding Eulerian numbers $\eulerian{n-3}{k-2}$.

In this paper we show that the minimal basis of independent $n$-particle partial amplitudes in the ${\rm N^{k-2}MHV}$ sector is indeed $\eulerian{n-3}{k-2}$-dimensional. The new expansion contains coefficients which are combinations of scalar and pseudo-scalar quantities. We also show that if one insists on having purely scalar coefficients, i.e. only functions of Mandelstam invariants, $s_{ab}=(k_a+k_b)^2$, then bases exist which are $\left(\eulerian{n-3}{k-2}+\eulerian{n-3}{n-k-2}\right)$-dimensional (or $2\eulerian{n-3}{k-2}$ since $\eulerian{n-3}{k-2}=\eulerian{n-3}{n-k-2}$) and are simultaneously valid for amplitudes in both the ${\rm N^{k-2}MHV}$ and ${\rm N^{n-k-2}MHV}$ sectors.

We provide several different ways of computing the coefficients by using a new set of objects we call {\it scalar blocks}. One of the most important properties of these new building blocks is that they provide a decomposition of a standard bi-adjoint scalar amplitude into ${\rm N^{k-2}MHV}$ sectors.

Amplitudes in a $U(N)\times U(\tilde N)$ bi-adjoint scalar theory can be double-flavor decomposed and therefore depend on two orderings $\alpha,\beta \in S_n$. The corresponding partial amplitude is usually denoted as $m_n(\alpha,\beta)$ \cite{Cachazo:2013iea}.

The ``${\rm N^{k-2}MHV}$ bi-adjoint" scalar blocks, which we denote as $m_{n,k}(\alpha,\beta)$, can be computed directly using the Cachazo-He-Yuan (CHY) formalism \cite{CHY, Cachazo:2013iea}. We also provide an alternative definition using Witten-RSV-like formulas \cite{witten, RSV}. In fact, the new scalar blocks are one of the missing items in the list of Witten-RSV formulas.

The new ${\rm N^{k-2}MHV}$ bi-adjoint objects can be decomposed into a scalar and a pseudo-scalar part\footnote{This means that the name {\it scalar blocks} seems to be an abuse of terminology. However, here we take {\it scalar} to refer to the amplitude the blocks construct when put together.}. Clearly, if a purely scalar quantity is desired then it can be obtained as follows
\begin{equation}
m_{n,k}^{\rm scalar}(\alpha,\beta) = \frac{1}{2}(m_{n,k}(\alpha,\beta) + m_{n,n-k}(\alpha,\beta)).
\end{equation}

Now we turn to a summary of our results. For Yang-Mills partial amplitudes in the ${\rm N^{k-2}MHV}$ sector we find relations of the form
\be
A_{n,k}^{\rm YM}(\gamma) = \sum_{i=1}^{\eulerian{n-3}{k-2}} F(\gamma,\alpha_i)A_{n,k}^{\rm YM}(\alpha_i )
\en
where
\be
F(\gamma,\alpha_i) = \sum_{j=1}^{\eulerian{n-3}{k-2}} m_{n,k}(\gamma, \beta_j)(m_{n,k})^{-1}(\beta_j,\alpha_i)
\en
while $\{ \alpha_i\}$ and $\{\beta_i\}$ are two sets, not necessarily the same, of $\eulerian{n-3}{k-2}$ permutations.

As mentioned above, the coefficients $F(\gamma,\alpha_i)$ in most ${\rm N^{k-2}MHV}$ sectors involve scalar and pseudo-scalar quantities. A notable exception is the helicity preserving sector where all coefficients are functions of only Mandelstam invariants.

If we impose the requirement that all coefficients be only functions of Mandelstam invariants for {\it all} sectors, then we find the decomposition
\be
A_{n,k}^{\rm YM}(\gamma) = \sum_{i=1}^{2\eulerian{n-3}{k-2}} F^{\rm scalar}(\gamma,\alpha_i)A_{n,k}^{\rm YM}(\alpha_i )
\en
where
\be
F^{\rm scalar}(\gamma,\alpha_i) = \sum_{j=1}^{2\eulerian{n-3}{k-2}} m^{\rm scalar}_{n,k}(\gamma, \beta_j)(m^{\rm scalar}_{n,k})^{-1}(\beta_j,\alpha_i)
\en
and now the sets of permutations contain $2\eulerian{n-3}{k-2}$ elements.

As a byproduct of the construction that leads to these two results we also find an intrinsically four-dimensional version of the KLT relation among gravity and gauge theory amplitudes. The new formula only involves $\eulerian{n-3}{k-2}$ partial amplitudes
\be
M^{\rm gravity}_{n,k} = \sum_{i,j=1}^{\eulerian{n-3}{k-2}}A^{\rm YM}_{n,k}(\alpha_i)\left(m^{-1}_{n,k}\right)(\alpha_i,\alpha_j)A^{\rm YM}_{n,k}(\alpha_j).
\en

The technique we use for these constructions is a simple generalization of the one given to derive the BCJ expansion of an amplitude using the CHY formalism in \cite{Cachazo:2013iea}. This argument, which only involves simple linear algebra, is reviewed in section 2. In section 3, we present the adaptation to four dimensions and the construction of the expansion in terms of the new basis. Section 3 also has the formal definition of scalar blocks, $m_{n,k}(\alpha,\beta )$, as given by their CHY formula as well as new KLT relations among gravity and gauge theory amplitudes. In section 4 we extend all previous constructions to the corresponding maximally supersymmetric versions. This is done by using the Witten-RSV formulations for gauge theory and gravity amplitudes. A Witten-RSV formula for the scalar blocks is also given in this section and plays a crucial role in the derivations. Section 5 is devoted to the study of some of the properties of the new scalar blocks. Most notably, we show that each scalar block is non-local but their sum over $k$-sectors is local. Section 6 reviews the BDFS quadratic relations and how their very large generalization provided by KLT orthogonality leads to a counting that shows the minimality of our $\eulerian{n-3}{k-2}$-dimensional basis. We end in section 7 with discussions of future directions.

\section{BCJ Basis and Its CHY Derivation}

Scattering amplitudes in $U(N)$ gauge theories can be written using a color decomposition \cite{Dixon:1996wi}. In its trace form, the color decomposition writes an amplitude as a linear combination of $(n-1)!$ color-ordered partial amplitudes. It is invariance under cyclic permutations that reduces the number of orderings from $n!$ down to $(n-1)!$.

These $(n-1)!$ partial amplitudes are not linearly independent. There are relations among them with constant coefficients known as the Kleiss-Kuijf (KK) relations \cite{KK} which reduce the elements in the basis down to $(n-2)!$. The construction of the KK basis is as follows. Using cyclic invariance it is possible to fix label $1$ as the first element. One possible KK basis is obtained by fixing label $n$ to be the last. This means that the KK relations must express any partial amplitude as a linear combination of elements in the basis. Explicitly,
\be\label{KKone}
A_n^{\rm YM}(1, \alpha , n , \beta ) =(-1)^{|\beta|}\sum_{\omega\in OP(\alpha, \beta^{\rm T})}A_n^{\rm YM}(1,\omega_2,\ldots ,\omega_{n-1},n)
\en
where $\alpha$ and $\beta$ are ordered subsets of labels so that the union of their elements form $\{2,3,\ldots ,n-1\}$. $OP(\alpha, \beta^{\rm T})$ are ``order preserving" permutations of all $n-2$ labels $\{2,3,\ldots ,n-1\}$. This means that when all labels in $\alpha$ (or $\beta$) are removed the leftover labels are in the order $\beta^{\rm T}$ (or $\alpha $). Here $\beta^{\rm T}$ is the reversed or transposed order of $\beta$.

When coefficients in the expansion are allowed to depend on Mandelstam invariants then there are more relations which are known as the Bern-Carrasco-Johansson (BCJ) relations \cite{Bern:2008qj}. These allow the dimension of the basis to be reduced from $(n-2)!$ down to $(n-3)!$. More explicitly, any partial amplitude in the KK basis can be written as
\be\label{BCJone}
A_n^{\rm YM}(1, \alpha, n-1,\beta , n ) =\sum_{\omega\in S_{n-3}}G(\alpha,\beta, \omega) A_n^{\rm YM}(1,\omega_2,\ldots ,\omega_{n-2},n-1,n)
\en
where $S_{n-3}$ denotes permutations of the labels in the set $\{ 2,3,\ldots ,n-2\}$ while the coefficients $G(\alpha,\beta, \omega)$ are functions of $s_{ab}=(k_a+k_b)^2$. These functions are known and we refer the reader to \cite{Bern:2008qj} for the explicit form.

Now we turn to the review of the derivation of the BCJ relations \eqref{BCJone} using the Cachazo-He-Yuan (CHY) representation of gluon amplitudes \cite{Cachazo:2015ksa}. Before turning to the derivation it is convenient to slightly generalize the form given above in \eqref{BCJone}.

In the following $\alpha$ denotes a general permutation of $n$ elements while ${\cal B}= \{\beta_I\}$ is a set of $(n-3)!$ permutations. The set ${\cal B}$ is not completely generic but we postpone the conditions it has to satisfy until more technology is introduced. The form of the BCJ relations we prove is 
\be
A_n^{\rm YM}(\alpha ) = \sum_{I=1}^{(n-3)!} F(\alpha,\beta_I)A_n^{\rm YM}(\beta_I)
\en
where again the coefficients $F$ depend on kinematic invariants of the form $s_{ab}=(k_a+k_b)^2$. One of the important facts about these relations is that the coefficients in the expansion do not depend on the polarization vectors chosen for the gluons. While the KK relations have an intuitive meaning from a lagrangian viewpoint, the BCJ relations do not seem to have a simple one. This means that there must be another formulation where the relations follow naturally. Below we review the CHY formulation and how it provides a natural origin for the BCJ relations based on simple linear algebra.

In the CHY formulation \cite{CHY}, tree amplitudes of $n$ massless particles are calculated by integrating over the moduli space of $n$ punctures on a Riemann sphere. The integral localizes to points in the moduli space which are solutions of the scattering equations:
\be
\sum_{b=1,b\ne a}^n \frac{s_{ab}}{\sigma_a-\sigma_b}=0\quad {\rm for}\quad a\in\{1,2,\ldots ,n\},
\en
where $s_{ab}$ are Mandelstam variables and $\sigma_a \in {\mathbb{C}}$ are the locations of the punctures. These equations have $(n-3)!$ solutions which is already a hint that there is some connection with the BCJ decomposition of amplitudes.

According to CHY, amplitudes in a variety of theories \cite{CHYdbi}, including Yang-Mills and Einstein gravity, can be written as
\be
\label{CHY}
A_n = \int d\mu_n \, {\cal I}_L {\cal I}_R=\sum_{I=1}^{(n-3)!}{\cal I}_L^{(I)} {\cal I}_R^{(I)} \frac{1}{\det' \Phi_I}.
\en
Here $d\mu_n$ is a meromorphic measure of integration over the moduli space (or $\sigma_a$'s modulo an ${\rm SL}(2,\mathbb{C})$ action), whose precise form is not important for our discussion. The only relevant fact is that the integral is a multidimensional contour integral and $d\mu_n$ contains poles at the solutions to the scattering equations thus defining the contour (for a precise definition see appendix A). The result is a sum of the product of the integrand, ${\cal I}_L {\cal I}_R$, and the jacobian, $\det' \Phi$, defined below, evaluated on all $(n-3)!$ solutions.

In all known theories to have a CHY representation the integrand splits into two sectors, the left integrand ${\cal I}_L$ and the right integrand ${\cal I}_R$ (each with half of the ${\rm SL}(2,\mathbb{C})$ weight to make the integral well-defined). This is the key to the existence of relations among amplitudes such as BCJ and the Kawai-Lewellen-Tye (KLT) relations \cite{KLT}.

The jacobian is computed using the reduced determinant of an $n\times n$ matrix
\begin{eqnarray}\Phi_{ab}=
\begin{cases}
\frac{s_{ab}}{\sigma_{ab}^2}, &a\ne b,
\cr -\sum_{c=1,c\ne a}^n \frac{s_{ac}}{\sigma_{ac}^2}, &a=b\end{cases}
\end{eqnarray}
which has co-rank three on the support of the solutions. Here we used the opportunity of introducing the shorthand notation $\sigma_{ab}=\sigma_a-\sigma_b$. The precise definition of the reduced determinant is
\be
{\rm det'}\Phi = \frac{\det \Phi^{abc}_{pqr}}{(\sigma_{ab}\sigma_{bc}\sigma_{ca})(\sigma_{pq}\sigma_{qr}\sigma_{rp})}
\en
where $\Phi^{abc}_{pqr}$ denotes the $(n-3)\times (n-3)$ matrix obtained from $\Phi$ by deleting rows $a,b,c$ and columns $p,r,q$. The fact that $\det'\Phi$ is independent of the choices is simple to prove and follows from arguments similar to those given by Hodges in his formulation of MHV gravity amplitudes \cite{Hodges} where he encountered a very similar matrix.

Let us now present the choices of integrands for the three theories relevant to this work: a biadjoint scalar, Yang-Mills and gravity.

The first theory is a bi-adjoint $\phi^3$ theory with flavor group $U(N)\times U(\tilde N)$. The double flavor-ordered decomposition of these amplitudes leads to partial amplitudes, $m_n(\alpha,\beta)$, specified by two permutations of $n$ labels denoted by $\alpha$ and $\beta$. The integrands ${\cal I}_L$ and ${\cal I}_R$ are Parke-Taylor-like factors:
\be\label{scalarI}
{\cal I}_L = \frac{1}{\sigma_{\alpha(1),\alpha(2)}\sigma_{\alpha(2),\alpha(3)}\cdots\sigma_{\alpha(n),\alpha(1)}},\quad {\cal I}_R= \frac{1}{\sigma_{\beta(1),\beta(2)}\sigma_{\beta(2),\beta(3)}\cdots\sigma_{\beta(n),\beta(1)}}.
\en
In order to keep the notation as compact as possible we introduce
\be
C_n(\gamma)=\frac{1}{(\gamma)}:=\frac{1}{\sigma_{\gamma(1),\gamma(2)}\sigma_{\gamma(2),\gamma(3)}\cdots\sigma_{\gamma(n),\gamma(1)}}.
\en

The next theory is Yang-Mills with color group $U(N)$. Here we have a single color decomposition and a partial amplitude is denoted by $A_n^{\rm YM}(\alpha)$ where $\alpha$ is the choice of permutation. The integrands are chosen to be
\be\label{YMI}
{\cal I}_L=C_n(\alpha),\quad {\cal I}_R={\rm Pf}'\Psi(\epsilon,k,\sigma).
\en
Here $\Psi(\epsilon,k,\sigma)$ is a $2n\times 2n$ antisymmetric matrix that contains the information about polarization vectors, $\epsilon_a$, for the gluons, momenta, $k_a$, and puncture locations $\sigma_a$. The matrix has co-rank 2 and hence ${\rm Pf}'\Psi(\epsilon,k,\sigma)$ denotes its reduced Pfaffian. The precise form of the matrix is not relevant for the purposes of this work but it is presented in appendix A for completeness.

Finally, graviton amplitudes in Einstein's theory (enriched to also contain a dilaton and a B-field) are denoted as $M_n$ and are computed using
\be\label{gravityI}
{\cal I}_L={\rm Pf}'\Psi(\epsilon,k,\sigma),
 \quad {\cal I}_R={\rm Pf}'\Psi(\tilde\epsilon,k,\sigma).
\en
Here we have allowed the polarization vectors on the two half-integrands to be different $\{\epsilon_a, \tilde\epsilon_a\}$. Given that each of the half-integrands is multi-linear in the polarization vectors, the combination of the two is multilinear in the rank-two tensors $\epsilon_{a,\mu\nu} = \epsilon_{a,\mu}\tilde\epsilon_{a,\nu}$. The tensor $\epsilon_{a,\mu\nu}$ is then taken to be the wave function of the $a^{\rm th}$ particle.

Now we are ready to present the derivation of the general form of relations among amplitudes that admit a CHY representation.

Let us use the label $I=1,2,\dots,(n-3)!$ to denote all solutions to the scattering equations and define $X(\sigma)_I$ as the value of a quantity $X(\sigma)$ that depends on the puncture locations, $\sigma_a$, at solution $I$.

Next, it is useful to define the matrices:
\be
D_{IJ}=\delta_{IJ}\frac{1}{(\det'\Phi)_I},~~ L_{I \alpha}=\frac{1}{(\alpha)_I}, ~~ {\rm and}~~ R_{I \beta}=\frac{1}{(\beta)_I}.
\en
The first matrix $D$ is an $(n-3)!\times (n-3)!$ matrix while the other two matrices, $L$ and $R$ are $(n-3)!\times n!$.

Thus a partial amplitude for the bi-adjoint $\phi^3$ theory is given as a component of a matrix obtained as the product
\be\label{oneT}
m_n(\alpha, \beta)=(L^{T} D R)_{\alpha \beta}=L_{\alpha I}D_{I J}R_{J \beta}
\en
with repeated indices summed over.

We would like to obtain a formula for the matrix $D$ from \eqref{oneT}. One immediate problem is that the matrices $L$ and $R$ are not square matrices. Moreover, \eqref{oneT} clearly implies that the $n!\times n!$ matrix $m_n$ has rank at most $(n-3)!$. The way out of this situation is to consider a $(n-3)!\times (n-3)!$ submatrix of $m_n$ of maximal rank. Such a submatrix is obtained by selecting two sets of permutations ${\cal A}$ and ${\cal B}$ with $(n-3)!$ elements each. Not all subsets are allowed because the Parke-Taylor-like factors $C_n(\gamma)$ also satisfy the KK relations \eqref{KKone} and hence the sets have to be chosen to be ``KK-independent".

Having selected ${\cal A}$ and ${\cal B}$ one can define square matrices $L_{\cal A}$ and $R_{\cal B}$ as the sub-matrices of $L$ and $R$ obtained from the corresponding sets of permutations. In the following we drop the subscripts and abuse the notation by calling the square matrices $L$ and $R$ again.

From \eqref{oneT} one can then obtain
\be
D=(L^T)^{-1}m R^{-1},~~ D^{-1}=R\; m^{-1} L^T.
\en

Given any amplitude that admits a CHY representation, $A^{\rm target}_n$, which is the ``target" for a decomposition,
\be
A^{\rm target}_n = \int d\mu_n\, {\cal I}_L\,{\cal I}_R = {\cal I}_L^T D\, {\cal I}_R
\en
one can write
\be
\label{LR}
A^{\rm target}_n = {\cal I}_L^T D D^{-1}D\, {\cal I}_R = ({\cal I}_L^T D R)\, m^{-1}(L^T D\, {\cal I}_R).
\en

The objects on the left, $({\cal I}_L^T D R)$, and on the right $(L^T D\, {\cal I}_R)$ are themselves amplitudes of other theories which contain a group factor $U(N)$ and therefore a color (or flavor) decomposition. Let us denote these amplitudes as $A^{\rm basis}_n(\alpha)=({\cal I}_L^T D R)_\alpha$ and $A^{\rm basis}_n(\beta)=(L^T D\, {\cal I}_R)_\beta$.

The general relation then takes the form
\be\label{genBCJ}
A^{\rm target}_n = \sum_{\alpha\in {\cal A},\beta\in{\cal B}}A^{\rm basis}_n(\alpha)(m^{-1})(\alpha,\beta)A^{\rm basis}_n(\beta).
\en

It is easy to see how these relations contain BCJ and KLT as particular cases. If ${\cal I}_L=C_n(\gamma),\, {\cal I}_R={\rm Pf}' \Psi(\kappa_a,\epsilon_a, \sigma_a)$, we have BCJ relations since $A^{\rm target}_n= A^{\rm YM}_n(\gamma)$, $A^{\rm basis}_n(\alpha) = m_n(\gamma,\alpha)$ and $A^{\rm basis}_n(\beta) = A^{\rm YM}_n(\beta)$. Explicitly,
\be
\label{BCJ}
A^{\rm YM}_n(\gamma)=\sum_{\beta\in {\cal B}}F(\gamma,\beta) A^{\rm YM}_n(\beta)
\en
with
\be
F(\gamma,\beta) = \sum_{\alpha\in {\cal A}} m_n(\gamma,\alpha)(m^{-1})(\alpha,\beta).
\en

Finally, if ${\cal I}_L={\cal I}_R={\rm Pf}' \Psi(\kappa_a,\epsilon_a, \sigma_a)$, then we obtain the KLT relations since $A^{\rm target}_n= M^{\rm gravity}_n$, $A^{\rm basis}_n(\alpha) = A^{\rm YM}(\alpha)$ and $A^{\rm basis}_n(\beta) = A^{\rm YM}(\beta)$. Explicitly,
\be\label{newKLT}
M^{\rm gravity}_n= \sum_{\alpha\in {\cal A},\beta\in{\cal B}} A^{\rm YM}_n(\beta)(m^{-1})(\beta, \alpha)A^{\rm YM}_n(\alpha).
\en

\section{Minimal Basis in Four Dimensions}

Four dimensions is very special from the point of view of the kinematics of massless particles. Many of the advances in the computation of scattering amplitudes of gluons have been triggered by the use of the spinor helicity formalism \cite{Elvang:2013cua}, which expresses a gluon momentum as
\be
k^a_\mu\sigma^\mu_{\alpha,\dot\alpha} = \lambda^a_\alpha \tilde\lambda^a_{\dot\alpha}.
\en
Spinor helicity variables, $\lambda^a_\alpha, \tilde\lambda^a_{\dot\alpha}$, make seemingly miraculous simplifications possible, such as those leading to the famous Parke-Taylor formula. Moreover, amplitudes become manifestly gauge invariant with the little group acting simply as a rescaling by a phase factor of the spinors. This leads to a separation of scattering amplitudes by sectors. Tree-level amplitudes become rational functions of the basic ${\rm SL}(2,\mathbb{C})$ invariants $\langle a,b\rangle$ and $[a,b]$. The standard definition of sectors is the following. An amplitude of gluons is in the ${\rm N^{k-2}MHV}$ sector if for any non-zero $t\in \mathbb{C}$,
\be
A_{n,k}^{\rm YM}(t\lambda_a ,t^{-1}\tilde\lambda_a) = t^{-2(n-2k)}A_{n,k}^{\rm YM}(\lambda_a ,\tilde\lambda_a).
\en
This definition coincides with having $k$ negative helicity gluons and $n-k$ positive helicity ones.

In 2003, Witten found that amplitudes in the ${\rm N^{k-2}MHV}$ sector are localized on curves of degree $k-1$ in twistor space \cite{witten}. Moreover, when interpreted as a string theory one has to integrate over maps from a punctured $\mathbb{CP}^1$ to twistor space of degree $k-1$.

Given that the CHY formulation unifies all the different sectors into a single one as required in arbitrary dimensions, one expects that the solutions to the scattering equations should branch out in four dimensions when spinor helicity variables are used.

Indeed, as shown in \cite{CHYthreedim}, the scattering equations are equivalent to asking a Lorentz vector of polynomials, $P^\mu(z)$, of degree $n-2$ to be null for any value of $z$. In four dimensions this means that the $2\times 2$ matrix $P_{\alpha,\dot\alpha}(z) = P^\mu(z)\sigma_{\mu,\alpha\dot\alpha}$ is singular, i.e. has vanishing determinant
\be
P_{1\dot 1}(z)P_{2\dot 2}(z) = P_{1\dot 2}(z)P_{2\dot 1}(z).
\en
Clearly, all the $2(n-2)$ roots on both sides must agree. This means that the roots of, say, $P_{1\dot 1}(z)$ must split between those of $P_{1\dot 2}(z)$ and $P_{2\dot 1}(z)$. Assuming that $d$ of them belong to the first factor and $n-2-d$ to the second, one can call the corresponding polynomials $\lambda_1(z)$ and $\tilde\lambda_{\dot 1}(z)$. Following this convention on notation one finds that for any $d\in \{0,1,\ldots ,n-2\}$
\be
P_{\alpha,\dot\alpha}(z) =\lambda_\alpha(z)\tilde\lambda_{\dot\alpha}(z)
\en
with ${\rm deg}\,\lambda_\alpha(z) =d$ and ${\rm deg}\,\tilde\lambda_\alpha(z) =n-d-2$. The scattering equations also require that
\be\label{link}
k^{a}_{\alpha\dot\alpha} = \frac{P_{\alpha\dot\alpha}(\sigma_a )}{\prod_{b\neq a}\sigma_{ab}}.
\en
This means that the cases with $d=0$ and $d=n-2$ do not lead to solutions for generic external momenta and must be discarded\footnote{E.g. when $d=0$, $\lambda_\alpha(\sigma)$ is a constant spinor, say $\rho_\alpha$. Using \eqref{link} one finds that $\rho^\alpha k^{a}_{\alpha\dot\alpha}=0$ for all $a$. For generic momenta this is only possible if $\rho_\alpha =0$.}. Therefore the solutions split into the following branches: $d=1,2,\ldots ,n-3$.

Also proven in \cite{CHYthreedim} is the act that the $(n-3)!$ solutions to the scattering equations separate so that $\eulerian{n-3}{d-1}$ solutions belong to the $d^{\rm th}$ branch. Therefore
\be
\eulerian{n-3}{0}+\eulerian{n-3}{1}+\dots+\eulerian{n-3}{n-4}=(n-3)!
\en
A very interesting property of the Yang-Mills and gravity integrands is that the sector definition obtained from the study of the little group has an intimate relation with that of branches of solutions. The relation is provided by the fact that the building block ${\rm Pf}'\Psi(\epsilon_1^{\mu},\epsilon_2^{\mu},\dots,\epsilon_n^{\mu},\sigma_1,\sigma_2,\dots,\sigma_n)$ with $k$ negative helicity polarization vectors and $n-k$ positive ones vanishes when evaluated on all solutions except for the ones corresponding to branch $d=k-1$.

Given this correspondence, we drop the distinction and talk about branches of solutions as sectors of solutions classified by $k$.

Working in the ${\rm N^{k-2}MHV}$ sector, an $n$-particle amplitude of gluons can then be written as
\be
A^{\rm YM}_{n,k}(\alpha)=\sum_{\{\sigma\}_I \in {\rm Sol(n,k)}}\frac{1}{(\det'\Phi)_I}\frac{({\rm Pf}' \Psi)_I}{(\alpha)_I}
\en
where the set of solutions in the $k^{\rm th}$-sector is denoted by Sol(n,k).

Following the same argument as in sector 2, it is natural to again introduce the matrices
\be
D_{k,IJ}=\delta_{IJ}\frac{1}{(\det'\Phi)_I},~~ L_{k,I \alpha}=\frac{1}{(\alpha)_I}, ~~ {\rm and}~~ R_{k,I \beta}=\frac{1}{(\beta)_I}.
\en
This time $D$ is an $\eulerian{n-3}{k-2}\times \eulerian{n-3}{k-2}$ matrix while the other two matrices, $L_k$ and $R_k$ are $\eulerian{n-3}{k-2}\times n!$.

Of course, the bi-adjoint $\phi^3$ theory, being a scalar theory, does not have any physical decomposition into helicity sectors. However, the separation of solutions induces a natural separation for this theory as well. We then define the {\it scalar blocks}
\be\label{oneT2}
m_{n,k}(\alpha, \beta):=(L_k^{T} D_k R_k)_{\alpha \beta}=\sum_{I,J\in {\rm Sol(n,k)}}L_{k,\alpha I}D_{k,I J}R_{k,J \beta}.
\en

The objects $m_{n,k}(\alpha, \beta)$ are new and section 5 is dedicated to the study of some of their properties. For the purposes of this section we only notice that they are rational functions of the basic ${\rm SL}(2,\mathbb{C})$ invariants $\langle a,b\rangle$ and $[a,b]$ such that when summed over $k=2,3,\dots, n-2$ give rise to the amplitude for bi-abjoint scalar theory, i.e.
\be
m_n(\alpha,\beta) = \sum_{k=2}^{n-2}m_{n,k}(\alpha, \beta).
\en 
This is in fact the motivation for their name. It is important to mention that $m_{n,k}(\alpha, \beta)$'s do not generically have definite parity, except for $k=n/2$ when they are scalars. This means that the name ``scalar blocks" seems to be an abuse of terminology. However, in this context ``scalar" refers to the amplitude one obtains by putting all the blocks together, i.e. $m_n(\alpha,\beta)$.

Continuing with the argument in section 2, one has to select a submatrix of size $\eulerian{n-3}{k-2}\times \eulerian{n-3}{k-2}$ from the $(n-3)!\times (n-3)!$ matrix of $m_{n,k}$. This is done by selecting two sets of permutations ${\cal A}_k$ and ${\cal B}_k$ with $\eulerian{n-3}{k-2}$ elements each. Again not all subsets are allowed because the Parke-Taylor-like factors $C_n(\gamma)$ in $m_{n,k}(\alpha,\beta)$ satisfy the KK relations \eqref{KKone} and hence the sets have to be chosen to be ``KK-independent".

We could denote the square matrices obtained as the obvious sub-matrices of $L_k$ and $R_k$ by $L_{{\cal A}_k}$ and $R_{{\cal B}_k}$. However, we again abuse notation and drop the subscripts of the sets and only keep the reference to the sector, i.e. $L_k$ and $R_k$.

The next steps are identical to those of section 2 so we will be brief. Using
\be
D_k=(L_k^T)^{-1}m_{n,k} R_k^{-1},~~ D_k^{-1}=R_k\; m_{n,k}^{-1} L_k^T.
\en
into
\be
A^{\rm target}_{n,k} =  {\cal I}_{L,k}^T D_k\, {\cal I}_{R,k}
\en
one can write
\be
\label{LR}
A^{\rm target}_{n,k} = ({\cal I}_{L,k}^T D_k R_k)m_{n,k}^{-1}(L_k^T D_k\, {\cal I}_{R,k}).
\en

The objects on the left, $({\cal I}_{L,k}^T D_k R_k)$, and on the right $(L^T_k D_k \,{\cal I}_{R,k})$ are themselves amplitudes of other theories in the $k^{\rm th}$-sector which contain a group factor $U(N)$ and therefore a color (or flavor) decomposition.

The general relation in the $k^{\rm th}$-sector is then
\be\label{kgen}
A^{\rm target}_{n,k} = \sum_{\alpha\in {{\cal A}_k},\beta\in {{\cal B}_k} }A^{\rm basis}_{n,k}(\alpha)(m_{n,k}^{-1})(\alpha,\beta)A^{\rm basis}_{n,k}(\beta).
\en
where the sums are over only $\eulerian{n-3}{k-2}$ elements instead of the $(n-3)!$ in the general relation \eqref{genBCJ} which is valid in arbitrary dimensions.

Applying the formula to the case when ${\cal I}_L=C_n(\gamma)$ and ${\cal I}_R={\rm Pf}' \Psi(\kappa_a,\epsilon_a, \sigma_a)$ one finds one of the main results of this work: a formula for expressing any partial amplitude in the ${\rm N^{k-2}MHV}$ sector in gauge theory as a linear combination of a basis with only $\eulerian{n-3}{k-2}$ elements,
\be
\label{kBCJ}
A^{\rm YM}_{n,k}(\gamma)=\sum_{\beta\in {\cal B}_k}F(\gamma,\beta) A^{\rm YM}_{n,k}(\beta)
\en
with
\be
F(\gamma,\beta) = \sum_{\alpha\in {\cal A}_k} m_{n,k}(\gamma,\alpha)(m_{n,k}^{-1})(\alpha,\beta).
\en

Likewise, this construction also provides a new KLT-like formula valid in four dimensions and which uses only $\eulerian{n-3}{k-2}$ gauge theory partial amplitudes on the left and right sectors:
\be\label{kKLT}
M^{\rm gravity}_{n,k}= \sum_{\alpha\in {\cal A}_k,\beta\in {\cal B}_k}A^{\rm YM}_{n,k}(\beta)(m_{n,k}^{-1})(\beta, \alpha)A^{\rm YM}_{n,k}(\alpha) .
\en

As explained in \cite{CHY}, general formulas of the form \eqref{kgen} can be used to connect many more theories than the ones we have discussed. Some examples include amplitudes of photons in Born-Infeld with Yang-Mills gluons and non-linear sigma model scalars. Our results in this section show that special relations valid in four dimensions also exist for all these theories as well.

\subsection*{Examples}

Let us use the general formulation to derive the three cases mentioned in the introduction. We start with the case of MHV or $\overline{\rm MHV}$ amplitudes and then proceed to the six-point NMHV amplitudes.

Scattering equations relevant to the MHV (or $\overline{\rm MHV}$) sector possess only one solution. The solution is
\be
\sigma_a-\sigma_b =r_ar_b \langle a~b\rangle.
\en
where the factors $r_a$ and $r_b$ turn out to drop out in the final formula (in the $\overline{\rm MHV}$ sector one simply exchanges $\langle a~b\rangle$ with $[a~b]$).

The evaluation of the biadjoint scalar amplitude on the MHV solution is straightforward and it leads to
\be\label{MHVm}
m_{n,2}(\alpha,\beta) = \frac{1}{\langle \alpha \rangle\langle \beta\rangle \det'\Phi_H}
\en
where the jacobian from the scattering equations, $\det'\Phi$, turns into Hodges' reduced determinant $\det'\Phi_H$. In order to make the connection precise let us review Hodges' definition \cite{Hodges}. Consider the $n\times n$ matrix
\begin{eqnarray}\Phi_{H,ab}=
\begin{cases}
\frac{[a,b]}{\langle a,b\rangle}, &a\ne b
\cr -\sum_{c=1,c\ne a}^n \frac{[a,c]}{\langle a,c\rangle}\frac{\langle x,c\rangle\langle y,c\rangle}{\langle x,a\rangle\langle y,a\rangle}, &a=b\end{cases}
\end{eqnarray}
where $x$ and $y$ represent two reference spinors. This matrix has co-rank three and its reduced determinant is
\be
{\rm det'}\Phi_H = \frac{\det \Phi^{abc}_{H,pqr}}{\langle a~b\rangle\langle b~c\rangle\langle c~a\rangle\,\langle p~q\rangle\langle q~r\rangle\langle r~p\rangle}.
\en
The definition of the remaining factors in \eqref{MHVm} is the following
\be
\langle \alpha \rangle \equiv \langle\alpha_1\alpha_2\rangle\langle\alpha_2\alpha_3\rangle\cdots \langle\alpha_n\alpha_1\rangle.
\en

Choosing the canonical ordering as the basis for amplitudes one has that our general formula \eqref{kBCJ} applied to $k=2$ implies
\be
A^{\rm YM}_{n,2}(\gamma )= m_{n,2}(\gamma,\beta)\frac{1}{m_{n,2}(\beta |1,2,\ldots ,n)} A^{\rm YM}_{n,2}(1,2,\ldots , n).
\en
Note that the answer should not depend on the choice of $\beta$ and indeed by using the explicit formula \eqref{MHVm} one finds the expected simple result \eqref{intro1}
\be
A^{\rm YM}_{n,2}(\gamma )=\frac{\langle 1,2,\ldots ,n \rangle}{\langle \gamma_1,\gamma_2,\ldots ,\gamma_n \rangle} A^{\rm YM}_{n,2}(1,2,\ldots , n).
\en
Note that nowhere in this derivation the explicit knowledge of the form of the MHV amplitudes was used. This kind of relations would have come as a surprise if we had not had the stunningly simple formulas provided by Parke and Taylor.

The analysis for the $\overline{\rm MHV}$ sector is entirely analogous.

Next, we show how any six-point NMHV partial amplitude can be written as a linear combination of a basis of only four partial amplitudes. Since NMHV is the helicity preserving sector when $n=6$, the coefficients of the expansion can all be expressed in terms of Mandelstam invariants $s_{ab}=(k_a+k_b)^2$.

We start by selecting two bases, ${\cal A}_3$ and ${\cal B}_3$, containing four permutations each. The subscript here indicates that we are in the $k=3$ sector. Given any general permutation of six labels $\gamma$ we find
\be
A^{\rm YM}_{6,3}(\gamma)=\sum_{\alpha \in {\cal A}_3}F(\gamma, \alpha) A^{\rm YM}_{6,3}(\alpha)
\en
with
\be
F(\gamma, \alpha)=\sum_{\beta\in {\cal B}_3}m_{6,3}(\gamma,\beta)(m_{6,3}^{-1})(\beta , \alpha ).
\en
Here $m_{6,3}$ is a $4\times 4$ matrix whose entries are the $k=3$ scalar blocks $m_{6,3}(\alpha, \beta)$ and $(m_{6,3}^{-1})$ is its inverse. In the next sections we provide several direct ways of directly computing these objects. Instead, here we use a special property valid for six particles:
\be\label{trick}
m_6(\alpha, \beta )  =  m_{6,2}(\alpha, \beta) + m_{6,3}(\alpha, \beta) + m_{6,4}(\alpha, \beta).
\en
Noting that $k=2$ is MHV and $k=4$ is $\overline{\rm MHV}$ one has
\be
m_{6,2}(\alpha, \beta) + m_{6,4}(\alpha, \beta) = \frac{1}{\langle \alpha \rangle\langle \beta\rangle \det'\Phi_H} + \frac{1}{[ \alpha ][ \beta] \det'\overline{\Phi}_H}.
\en
This combination is clearly parity invariant and therefore a function of only Mandelstam invariants. Using \eqref{trick} the scalar blocks of interest are 
\be
m_{6,3}(\alpha, \beta) = m_6(\alpha, \beta ) -\left(\frac{1}{\langle \alpha \rangle\langle \beta\rangle \det'\Phi_H} + \frac{1}{[ \alpha ][ \beta] \det'\overline{\Phi}_H}\right)
\en
where $m_6(\alpha, \beta )$ is the standard bi-adjoint $\phi^3$ double partial amplitude.

Before finishing this section, we point out a convenient choice of permutations. Let the sets be
\bes
\label{basis}
{\cal A}_3 &=& \{(124356),(142356),(143256),(134256)\},\tr
{\cal B}_3 &=& \{ (153462),(154362),(152463),(154263)\}.
\ens
The reason for this choice is that the matrix for the standard bi-adjoint scalar theory, $m_6(\alpha,\beta)$, takes a particulary simple form
\be
m_6 = \left(
        \begin{array}{cccc}
          \frac{s_{34}+s_{35}}{s_{12}s_{34}s_{35}s_{126}} & \frac{-1}{s_{12}s_{34}s_{126}} & 0 & 0 \\
          0 & 0 & 0 & 0 \\
          0 & 0 & 0 & 0 \\
          0 & 0 &  \frac{s_{24}+s_{25}}{s_{13}s_{24}s_{25}s_{136}} & \frac{-1}{s_{13}s_{24}s_{136}} \\
        \end{array}
      \right).
\en
Here we have used the shorthand notation $s_{abc}=s_{ab}+s_{bc}+s_{ca}$. A calculation of the matrix elements in $m_6$ using trivalent graphs is illustrated in appendix B.

\section{SUSY and The Witten-RSV Formulation}

The derivation of the minimal basis of four-dimensional partial amplitudes given in the previous section was based on the CHY formulation of scattering amplitudes. While the CHY formulation allows us to find a general relation, \eqref{kBCJ}, which contains, as particular cases, the four dimensional analogs of BCJ and KLT, it also implies many other relations among other classes of theories. A limitation of the CHY derivation is that all the analysis is done for bosonic fields. In this section we show that by using the Witten-RSV formulation \cite{witten, RSV} it is possible to prove that the formulas for the minimal basis of Yang-Mills amplitudes, \eqref{kBCJ}, and the new KLT formula for gravity, \eqref{kKLT}, are both valid for their corresponding maximally supersymmetric versions.

The most convenient starting point is the manifestly parity invariant version of the Witten-RSV formulas \cite{Witten:2004cp, Cachazo:2013zc}. Let us review some of the relevant details of the formulation before using it.

Amplitudes in ${\cal N}=4$ super Yang-Mills in the canonical order are given by
\be
{\mathfrak{A}}^{\rm SYM}_{n,k}(1,2,\ldots ,n) = \int d\Omega^{B}_{n,k}\, C_n(1,2,\ldots ,n)\int d^2\Omega^F_{n,k}R(\lambda)R(\tilde{\lambda})
\en
where $d\Omega^{B}_{n,k}$ and $d^2\Omega^{F}_{n,k}$ are bosonic and fermionic measures respectively. The bosonic measure contains the familiar integration over $n$ puncture locations $\sigma_a$ and is defined in detail below. The precise form of the fermionic measure is not relevant for our discussion and can be found in \cite{Cachazo:2013zc}. $C_n(1,2,\ldots ,n)$ is $1/(\sigma_{12}\sigma_{23}\cdots \sigma_{n1})$ as usual. The factors $R(\lambda)$ and $R(\tilde{\lambda})$ are resultants of polynomials as explained below. The reader familiar with the literature might recall that in Yang-Mills amplitudes the resultants are supposed to cancel. This is indeed true and they cancel with similar factors in the measure $d\Omega^{B}_{n,k}$. However, it is instructive to exhibit them explicitly so that the two parts of the integrand have the correct ${\rm SL}(2,\mathbb{C})$ transformations as a left, ${\cal I}_L$, and right, ${\cal I}_R$, integrands in previous sections.

The reason for changing the notation of gauge theory amplitudes from $A^{\rm SYM}_{n,k}$ to ${\mathfrak{A}}^{\rm SYM}_{n,k}$ is that the new amplitudes contain momentum conserving delta functions while the old ones do not.

Next we present the equivalent formulation for ${\cal N}=8$ supergravity.
\be
{\mathfrak{M}}_{n,k}^{\rm sugra} = \! \int \! d\Omega^{B}_{n,k}\,\int d^2\Omega^F_{n,k}R(\lambda)R(\tilde{\lambda})\,\int d^2\widehat\Omega^F_{n,k}R(\lambda)R(\tilde{\lambda}).
\en
This formula can be obtained from the Yang-Mills one by simply replacing the Parke-Taylor like factor by another copy of the supersymmetry integrand. This is why the amount of supersymmetry is doubled.

Finally, it is obvious now what formula should describe the biadjoint scalar blocks
\be\label{mWRSV}
{\mathfrak{m}}_{n,k}(\alpha, \beta) = \int d\Omega^{B}_{n,k}\, C_n(\alpha)\, C_n(\beta ).
\en
Once again, the new notation ${\mathfrak{m}}_{n,k}(\alpha, \beta)$ indicates that this object contains a momentum conserving delta function while $m_n(\alpha,\beta )$ does not.

This is our second formula for computing scalar blocks in the ${\rm N^{k-2}MHV}$ sector. Let us write down the ingredients in this formula in detail. The measure is defined as
\bes \label{spinor}
d\Omega^{B}_{n,k} & = & \frac{1}{{\rm vol}\,{\rm SL}(2,\mathbb{C})\times \mathbb{C}^*}\prod_{a=1}^nd\sigma_a dt_a d{\tilde t}_a\delta\left(t_a{\tilde t}_a-\frac{1}{\prod_{b\neq a}\sigma_{ab}}\right)\times \tr & & \prod_{r=0}^d d^2\rho_r\delta^2\left(\lambda_a-t_a\lambda(\sigma_a)\right)\prod_{r=0}^{\tilde d} d^2\tilde\rho_r\delta^2\left(\tilde\lambda_a-{\tilde t}_a\tilde\lambda(\sigma_a)\right)\times \tr & & \frac{1}{R(\lambda)R(\tilde{\lambda})}
\ens
where $d=k-1$, $\tilde d =n-k-1$ and the two polynomials spinors are given by
\be
\lambda(\sigma) = \sum_{r=0}^d\rho_r\sigma^r, \quad \tilde\lambda(\sigma) = \sum_{r=0}^{\tilde d}\tilde\rho_r\sigma^r.
\en
Finally, $R(\lambda)$ is the resultant of the polynomials $\{\sum_{i=0}^d\rho_{r,1}\sigma^r,\sum_{r=0}^d\rho_{r,2}\sigma^r \}$ while $R(\tilde\lambda)$ is that of $\{\sum_{r=0}^{\tilde d}\tilde\rho_{r,1}\sigma^r,\sum_{r=0}^{\tilde d}\tilde\rho_{r,2}\sigma^r \}$.

The derivation of the relation among gauge theory amplitudes is now straightforward and parallels that using the CHY representation. Each of the formulas presented above is evaluated as a sum over $\eulerian{n-3}{k-2}$ solutions and therefore can be written as
\be
\delta^4\left(k_1+k_2+\ldots +k_n\right)\!\!\!\sum_{\{\sigma\}_i \in {\rm Sol(n,k)}}({\cal I}_L)_i D_{k,ii}({\cal I}_R)_i.
\en
In the case of the scalar blocks it is convenient to always remove the momentum conserving delta function which then gives rise to the same definition we obtained using the CHY representation.

Using that the scalar blocks are the same as in the previous section we can immediately write
\be
\label{kSBCJ}
{\mathfrak{A}}^{\rm SYM}_{n,k}(\gamma)=\sum_{\beta\in {\cal B}_k}F(\gamma,\beta) {\mathfrak{A}}^{\rm SYM}_{n,k}(\beta)
\en
with
\be
F(\gamma,\beta) = \sum_{\alpha\in {\cal A}_k} m_{n,k}(\gamma,\alpha)(m_{n,k}^{-1})(\alpha,\beta).
\en
Likewise, this construction also provides a new KLT formula for ${\cal N}=8$ supergravity and which uses $\eulerian{n-3}{k-2}$ gauge theory partial amplitudes on the left and right sectors:
\be\label{kSKLT}
{\mathfrak{M}}^{\rm sugra}_{n,k}= \sum_{\alpha\in {\cal A}_k,\beta\in {\cal B}_k}A^{\rm SYM}_{n,k}(\beta)(m_{n,k}^{-1})(\beta, \alpha){\mathfrak{A}}^{\rm SYM}_{n,k}(\alpha).
\en
Note that on the right hand side one of the super-Yang-Mills amplitudes, ${\mathfrak{A}}^{\rm SYM}_{n,k}(\alpha)$, carries a momentum conserving delta function while the other, $A^{\rm SYM}_{n,k}(\beta)$, does not.

\section{Scalar Blocks}

This section is devoted to the study of the new objects found this work; the scalar blocks. In sections 2 and 3 we introduced two definitions for the same object. On the one hand there is the CHY formulation \eqref{oneT2}
\be\label{mnk}
m_{n,k}(\alpha,\beta) = \sum_{I\in {\rm Sol(n,k)}}\frac{1}{{\rm det}'\Phi_I}\frac{1}{(\alpha_1,\alpha_2,\ldots ,\alpha_n)_I}\,\frac{1}{(\beta_1,\beta_2,\ldots ,\beta_n)_I}
\en
while on the other is the Witten-RSV formulation \eqref{mWRSV}
\be\label{ScalarBlock}
{\mathfrak{m}}_{n,k}(\alpha, \beta) = \int\frac{d\widehat{\Omega}^{B}_{n,k}}{R(\lambda)R(\tilde\lambda )}\,\frac{1}{(\alpha_1,\alpha_2,\ldots ,\alpha_n)}\, \frac{1}{(\beta_1,\beta_2,\ldots ,\beta_n)}.
\en
Here we have chosen to exhibit the factors $R(\lambda)$ and $R(\tilde\lambda )$ explicitly as they are important for the following. This means that the old measure in \eqref{mWRSV} and the new are related by $d\widehat{\Omega}^{B}_{n,k}=R(\lambda)R(\tilde\lambda )d\Omega^{B}_{n,k}$.

From the first definition \eqref{mnk} it is obvious that the sum over all values of $k$ gives rise to the standard bi-adjoint scalar amplitude
\be\label{newS}
m_n(\alpha , \beta ) = \sum_{k=2}^{n-2}m_{n,k}(\alpha ,\beta ).
\en
When $m_{n,k}(\alpha ,\beta )$ are written using the Witten-RSV formula, \eqref{newS} actually gives a new formulation for the scalar amplitude thus giving one new item in the four-dimensional dictionary started by Witten in 2003.

In either formulation it is clear that although the blocks $m_{n,k}(\alpha ,\beta )$ do not carry helicity, they are neither scalar nor pseudo-scalars. In fact, under a parity transformation $m_{n,k}(\alpha ,\beta )$ and $m_{n,n-k}(\alpha ,\beta )$ are exchanged. This motivates us to define the natural scalar combination
\be\label{scalarblock}
m_{n,k}^{\rm scalar}(\alpha ,\beta ) =\frac{1}{2}\left( m_{n,k}(\alpha ,\beta )+ m_{n,n-k}(\alpha ,\beta )\right).
\en
This is parity invariant combination of scalar blocks is only a function of Mandelstam invariants.

The general formula for relations among amplitudes found in section 2 can easily be modified to use the new combinations of scalar blocks and give rise to expansions whose coefficients are purely scalars. The price to pay for using only Mandelstam invariants is that the dimension of the basis goes from $\eulerian{n-3}{k-2}$ to $\eulerian{n-3}{k-2}+\eulerian{n-3}{n-k-2}$. It is well-known that $\eulerian{n-3}{k-2}=\eulerian{n-3}{n-k-2}$ and hence the basis only doubles its size. The formulas are obtained by selecting sets of permutations of the appropriate size which are ``KK independent" as described in section 2. The final formula, which is valid for amplitudes in both the $k$ and $n-k$ sectors, reads
\be
\label{PIBCJ}
A^{\rm YM}_{n,(k,n-k)}(\gamma)=\sum_{\beta\in {\cal B}_{(k,n-k)}}F_{n,(k,n-k)}^{\rm scalar}(\gamma,\beta) A^{\rm YM}_{n,(k,n-k)}(\beta)
\en
with
\be
F_{n,(k,n-k)}^{\rm scalar}(\gamma,\beta) = \sum_{\alpha\in {\cal A}_{(k,n-k)}} m^{\rm scalar}_{n,k}(\gamma,\alpha)(m^{\rm scalar}_{n,k})^{-1}(\alpha,\beta).
\en

We now go back to $m_{n,k}(\alpha,\beta)$ and in the rest of this section study its structural properties.

The simplest observation is that $m_{n,k}(\alpha,\beta)$ is a rational function of invariants $\langle a~b\rangle$ and $[a~b]$ which has helicity zero in all particles. As a rational function, the first natural question is its pole structure. From its definition it is clear that poles of $m_{n,k}(\alpha,\beta)$ can only have two possible origins. The first is related to kinematic invariants that lead to two or more puncture locations $\sigma_a$ to coincide. The second is from kinematic invariants that lead to degenerations of the maps $\lambda(\sigma)$ and $\tilde\lambda(\sigma )$ giving rise to vanishing resultants, $R(\lambda)$ or $R(\tilde\lambda)$. The former kind of poles has been well-studied in the literature as it is the only kind of poles of the standard biadjoint scalar amplitude and correspond to physical factorization channels. The second class of poles is new and corresponds to unphysical poles. In fact, one expects two different kinds of unphysical poles; one for each resultant. This means that
\be
m_{n,k}(\alpha,\beta) = \frac{N_{n,k}(\alpha,\beta)}{Q_{n,k}{\tilde Q}_{n,k}\prod_{i\in P} s_i}
\en
where $P$ is the set of physical poles of the biadjoint scalar amplitude $m_{n}(\alpha,\beta)$ while $Q_{n,k}$ and ${\tilde Q}_{n,k}$ are the two sets of unphysical poles (it is easy to see that the only exceptions are $Q_{n,2}$ and $\tilde{Q}_{n,n-2}$ which do not give rise to any unphysical poles).

It is clear that the unphysical poles must be shared among different scalar blocks with different values of $k$ since they must completely cancel in the combination that leads to the physical biadjoint scalar amplitude \eqref{newS}. We have found that the pattern of cancellations is the simplest possible one: {\it unphysical poles cancel in pairs between neighbors thus forming a linear chain.} More explicitly, writing \eqref{newS} in expanded form
\be
m_n(\alpha , \beta ) = m_{n,2}(\alpha ,\beta )+m_{n,3}(\alpha ,\beta )+m_{n,4}(\alpha ,\beta )+\ldots.
\en
one finds that in the three terms shown below
\be
\left( \ldots + \frac{N_{n,k-1}(\alpha,\beta)}{Q_{n,k-1}{\tilde Q}_{n,k-1}} +\frac{N_{n,k}(\alpha,\beta)}{Q_{n,k}{\tilde Q}_{n,k}}+\frac{N_{n,k+1}(\alpha,\beta)}{Q_{n,k+1}{\tilde Q}_{n,k+1}}\ldots \right)\frac{1}{\prod_{i\in P} s_i}
\en
the poles $Q_{n,k}$ and ${\tilde Q}_{n,k}$ are absent. The reason this is even possible in the first place is that ${\tilde Q}_{n,k-1}=Q_{n,k}$ for $k=3,4,\dots, n-2$. Let us now prove this last statement. 

From \eqref{ScalarBlock} we know that $Q_{n,k}$ comes from the zeroes of  the resultant $R(\lambda^{(k-1)})$ for polynomials $\lambda_{1}^{(k-1)}(\sigma)$ and $\lambda_{2}^{(k-1)}(\sigma)$ and $\tilde{Q}_{n,k}$ from $R(\tilde{\lambda}^{(n-k-1)})$, where we temporarily denote $\lambda_{\alpha}^{(d)}(\sigma)$ and $\tilde{\lambda}_{\dot{\alpha}}^{(d)}(\sigma)$ as polynomials of degree $d$ for later convenience. Thus when $R(\lambda^{(k-1)})=0$, from \eqref{spinor} we have
\be\label{resultant}
\lambda_{\alpha}^a=t'_a (\sigma_a-\sigma_*)\lambda_{\alpha}^{(k-2)}(\sigma_a), ~~ \tilde{\lambda}_{\dot \alpha}^a=\tilde{t}_a \tilde\lambda_{\dot{\alpha}}^{(n-k-1)}(\sigma_a).
\en
Using the $\mathbb{C}^*$ gauge freedom of $t_a$ and $\tilde{t}_a$ it is possible to rewrite the above as:
\be\label{resultant2}
\lambda_{\alpha}^a=t_a \lambda_{\alpha}^{(k-2)}(\sigma_a),~~ \tilde{\lambda}_{\dot \alpha}^a=\tilde{t}'_a (\sigma_a-\sigma_*) \tilde\lambda_{\dot{\alpha}}^{(n-k-1)}(\sigma_a)
\en
with
\be
t_a=t'_a (\sigma_a-\sigma_*), ~~ \tilde{t}_a=\tilde{t}'_a (\sigma_a-\sigma_*).
\en
Consequently, $R(\tilde{\lambda}^{(n-k'-1)})=0$ where $k'=k-1$. Since the argument above can be reversed, we now have:
\be
R(\lambda^{(k-1)})=0\, \Leftrightarrow \, R(\tilde{\lambda}^{(n-k'-1)})=0.
\en
Thus we conclude there is a one-to-one unphysical-pole correspondence between ${\tilde Q}_{n,k-1}$ and $Q_{n,k}$ for $k=3,4,\dots, n-2$.

\subsection*{Example: $(k,n-k)=(2,n-2)$}

In this part of the section we illustrate the use of the scalar version of the formula, \eqref{PIBCJ}, expressing partial amplitude in terms of a basis of dimension $2\eulerian{n-3}{k-2}$. We concentrate on the case with $k=2$.
From eq. \eqref{mnk} and eq. \eqref{scalarblock} we write the scalar block for the MHV+${\overline{\rm MHV}}$ sector:
\be
m_{n,2}^{\rm scalar}(\alpha,\beta)=\frac{1}{2}\Big( \frac{1}{\langle \alpha \rangle\langle \beta\rangle \det'\Phi_H} + \frac{1}{[ \alpha ][ \beta] \det'\overline{\Phi}_H}\Big)
\en
From this expression, we write eq. \eqref{PIBCJ} as
\be\label{MHV}
A^{\rm YM}_{n,(2,n-2)}(\gamma)=\sum_{i=1}^2F_{n,(2,n-2)}^{\rm scalar}(\gamma,\beta_i) A^{\rm YM}_{n,(2,n-2)}(\beta_i)
\en
where $F^{\rm scalar}_{n,(2,n-2)}(\gamma,\beta_i)$ depends only on Mandelstam variables because of parity invariance. The coefficients  $F^{\rm scalar}_{n,(2,n-2)}(\gamma,\beta_i)$ can be explicitly written down
\be \label{mhvmatrix}
 F^{\rm scalar}_{n,(2,n-2)}(\gamma,\beta_i)=\frac{\la \beta_i \ra [\beta_i]}{\la \gamma \ra [\gamma]} \frac{\la \beta_{i+1} \ra [\gamma]-\la \gamma \ra [\beta_{i+1}]}{\la \beta_{i+1} \ra [\beta_i]-\la \beta_i \ra [\beta_{i+1}]},\quad i=1,2
\en
where all indices of permutations are understood modulo 2, i.e., $\beta_3 \equiv \beta_1$. 

Restricting to the six-particle case, we are able to express \eqref{mhvmatrix} explicitly in terms of Mandelstam variables by choosing the basis of permutations to be ${\cal B}_{(2,4)} = \{ (1,2,3,4,5,6),(1,2,3,4,6,5) \}$. Standard BCJ relations can be used to fix three labels, say, $1,2,3$, so we only need to consider the remaining four orderings  which are not contained in our basis. For example, choosing $\gamma= (1,2,3,5,4,6)$ eq. \eqref{mhvmatrix} becomes:
\bes
F_{6,(2,4)}^{\rm scalar}(\gamma,\beta_1)= \frac{s_{45}S_{3615}^{4615}-s_{35}S_{4615}^{4615}}{s_{35}S_{4615}^{4615}},~~
F_{6,(2,4)}^{\rm scalar}(\gamma,\beta_2)=-\frac{s_{15}S_{3546}^{4615}}{s_{35}S_{4615}^{4615}}
\ens
with
\be
S_{pqrt}^{abcd} := {\rm det}\left(
                              \begin{array}{cccc}
                                s_{ap} & s_{aq} & s_{ar} & s_{at} \\
                                s_{bp} & s_{bq} & s_{br} & s_{bt} \\
                                s_{cp} & s_{cq} & s_{cr} & s_{ct} \\
                                s_{dp} & s_{dq} & s_{dr} & s_{dt} \\
                              \end{array}
                            \right).
\en
The other three permutations are computed in a completely analogous way.

\section{Connection to Quadratic Relations}

In 2011, Bjerrum-Bohr, Damgaard, Feng and Sondergaard \cite{BDFS} asked the following question. If KLT relations are written as
\be
M_{n,k}^{\rm gravity} = \sum_{i,j=1}^{(n-3)!}A_{n,k}^{\rm YM}(\alpha_i) {\cal S}(\alpha_i,\beta_j)A_{n,k}^{\rm YM}(\beta_j)
\en
with ${\cal S}(\alpha,\beta)$ the well-known momentum kernel \cite{Bern:2008qj}, then what would happen if the helicity sector of the Yang-Mills amplitudes on the left, $A_{n,k}^{\rm YM}(\alpha)$, is chosen to be different from the helicity sector of amplitudes on the right, $A_{n,k'}^{\rm YM}(\beta)$? The answer turns out to be {\it quadratic relations} among Yang-Mills amplitudes. 

More explicitly, when $k\neq k'$ then
\be
\label{quadratic}
\sum_{i,j=1}^{(n-3)!}A_{n,k}^{\rm YM}(\alpha_i) {\cal S}(\alpha_i,\beta_j) A_{n,k'}^{\rm YM}(\beta_j)=0.
\en
Recall that $\{\alpha_i\}$ and $\{\beta_i\}$ are two sets, not necessarily equal, of $(n-3)!$ KK-independent permutations. 

The presence of new relations among partial amplitudes led to the idea that the $(n-3)!$ dimensional BCJ basis might not be minimal in a four-dimensional space-time \cite{BDFS, He-Feng}.

In previous sections we constructed a basis of partial amplitudes in four dimensions which is indeed smaller than the BCJ basis. The natural question is the relation of our construction to the quadratic relations.

One of the first explicit proposals for reducing the basis uses the quadratic relations with fixed $k$ different from $2$ or $n-2$ and $k'\in \{2,n-2\}$ \cite{BDFS, He-Feng}. This gives rise to two simple equations among the $(n-3)!$ partial amplitudes $A_{n,k}^{\rm YM}(\alpha)$:
\bes
\sum_{i=1}^{(n-3)!}A_{n,k}^{\rm YM}(\alpha_i) \sum_{j=1}^{(n-3)!}\frac{{\cal S}(\alpha_i,\beta_j)}{\langle\beta_j\rangle} & = & 0, \tr \sum_{i=1}^{(n-3)!}A_{n,k}^{\rm YM}(\alpha_i) \sum_{j=1}^{(n-3)!}\frac{{\cal S}(\alpha_i,\beta_j)}{[\beta_j]} & = & 0.
\ens
Using these two equations one can reduce the basis down to $(n-3)!-2$. 

When $n=6$ the basis has dimension four. In fact, the linear relations can be simplified and written down explicitly \cite{He-Feng}
\bes\label{sixrel}
[2|3+4+5|6\rangle[3|4+5|6\rangle[4|5|6\rangle A_{6,3}^{\rm YM}(1,2,3,4,5,6) + {\cal P}(234)  & = & 0, \tr
\langle 2|3+4+5|6]\langle 3|4+5|6]\langle 4|5|6] A_{6,3}^{\rm YM}(1,2,3,4,5,6) + {\cal P}(234)  & = & 0.
\ens

Noting that from our construction we also have $\eulerian{n-3}{k-2} = \eulerian{3}{1} =4$, it must be that our basis coincides with the one
obtained from using the relations \eqref{sixrel}. We have checked that this is indeed the case.

For general values of $n$ and $k$ the quadratic relations are not enough to lower $(n-3)!$ down to $\eulerian{n-3}{k-2}$. One way to see this is by finding an upper bound on the number of possible quadratic relations. This gives the maximum number of elements that can be removed from the $(n-3)!$-dimensional BCJ basis. 

Let us write the relations obtained by using $k'=2,3,\ldots ,\hat{k},\ldots ,n-2$ as
\bes\label{cons}
\sum_{i=1}^{(n-3)!}A_{n,k}^{\rm YM}(\alpha_i) \left(\sum_{j=1}^{(n-3)!}{\cal S}(\alpha_i,\beta_j)A_{n,k'}^{\rm YM}(\beta_j) \right)= 0.
\ens
For each of the $n-4$ values of $k'$ there are $\binomial{n}{k'}$ possibilities for the choices of the $k'$ negative helicity gluons in $A_{n,k'}^{\rm YM}(\beta)$. Of course, this is an upper bound since Ward identities can drastically reduce this number, e.g., when $k'=2$ one goes from $\binomial{n}{2}$ down to $1$. This means that there are at most $\sum_{k'\neq k}\binomial{n}{k'}$ relations\footnote{One could also include gluinos and scalars in the $k'$ sector but the number of relations is still polynomial in $n$.}. Therefore the reduced basis cannot be smaller than $(n-3)!-\sum_{k'\neq k}\binomial{n}{k'}$. 

Since the lower bound $(n-3)!-\sum_{k'\neq k}\binomial{n}{k'}$ is much larger than $\eulerian{n-3}{k-2}$ for $n>8$ and any $k$, the natural question is where the remaining relations that lead to our $\eulerian{n-3}{k-2}$-dimensional basis come from.

The answer to this question comes from a property discovered by Geyer and one of the authors called KLT orthogonality \cite{Cachazo and Geyer}. The property was proven in \cite{CHYthreedim} and it states that
\bes
\sum_{i=1}^{(n-3)!}\frac{1}{(\alpha_i )_J} \left(\sum_{j=1}^{(n-3)!}{\cal S}(\alpha_i,\beta_j)\frac{1}{(\beta_j)_I} \right)= 0
\ens
for any two different solutions, $I$ and $J$, of the scattering equations. Recall the meaning of the notation $(\alpha)=\sigma_{\alpha_1\alpha_2}\sigma_{\alpha_2\alpha_3}\cdots \sigma_{\alpha_{n}\alpha_1}$. This means the objects $1/(\alpha)$ evaluated on a given solution, $J$, satisfie $(n-3)!-1$ relations.

In order to find the implications of KLT orthogonality to relations among partial amplitudes let us write
\be
A_{n,k'}^{\rm YM}(\beta) = \sum_{\{\sigma \}_I\in {\rm Sol}(n,k')}\frac{1}{(\beta)_I}D_{II}(\int d^2\Omega^F_{n,k}R_{n,k'}(\lambda)R_{n,n-2-k'}(\tilde{\lambda}))_I.
\en
Note that now the relation \eqref{cons} follows from $\eulerian{n-3}{k'-2}$ more refined relations:
\bes
\sum_{i=1}^{(n-3)!}A_{n,k}^{\rm YM}(\alpha_i) \left(\sum_{j=1}^{(n-3)!}{\cal S}(\alpha_i,\beta_j)\frac{1}{(\beta_j)_I} \right)= 0,
\ens
i.e., one for every $I\in {\rm Sol}(n,k')$.

It is important to notice that the factors containing the helicity information of the particles, i.e. the fermionic measure, in $D_{II}(\int d^2\Omega^F_{n,k}R_{n,k'}(\lambda)R_{n,n-2-k'}(\tilde{\lambda}))_I$ drop out of the relations since they are permutation invariant in the labels.

The refined versions now provide $\eulerian{n-3}{k'-2}$ relations for each $k'$ and hence the $(n-3)!$ basis is reduced to
\be
(n-3)!-\sum_{k'\neq k}\eulerian{n-3}{k'-2}=\eulerian{n-3}{k-2}
\en
which is in exact agreement with our proposal of a $\eulerian{n-3}{k-2}$ dimensional basis! Moreover, since the helicity information of the $k'$ pieces drops out it is not possible to obtain extra relations and this is the reason why the basis we have found is the minimal basis in four dimensions.

\section{Discussions}

The $(n-3)!$-dimensional BCJ basis of Yang-Mills partial n-particle amplitudes \cite{Bern:2008qj}, introduced in 2008, is a major improvement compared to the $(n-2)!$-dimensional KK basis \cite{KK}. The price for the improvement on the KK basis is to allow the coefficients in the expansion of an arbitrary amplitude in terms of the basis to depend on Mandelstam invariants while the KK basis only uses constant coefficients. The BCJ basis is believed to be the minimal basis when the dimension of spacetime is generic. Already in 2011, the BDFS quadratic relations for partial amplitudes in four dimensions implied the existence of smaller basis \cite{BDFS,He-Feng}. In this work, we have found a basis of dimension $\eulerian{n-3}{k-2}$ for ${\rm N^{k-2}MHV}$ Yang-Mills partial n-particle amplitudes. We have also shown that this is in fact the minimal basis in four dimensions. What makes four dimensions special is the decomposition into different R-charge sectors and the spinor helicity formalism which induces a separation into branches of the solutions to the scattering equations.

The new basis has coefficients which carry zero helicity but are not necessarily parity invariant. In spinor-helicity terminology, the coefficients of the new basis are rational functions of the basic ${\rm SL}(2,\mathbb{C})$ invariants $\langle a~b\rangle$ and $[a~b]$ which are invariant under rescalings of the form $(\lambda_a,\tilde\lambda_a) \to (t_a\lambda_a, t^{-1}_a\tilde\lambda_a)$ for any non-zero complex numbers $t_a$.

We also showed that if one insists on a basis such that the coefficients are only scalars, i.e., only functions of Maldelstam invariants $s_{ab}=(k_a+k_b)^2$ then this is also possible but at the expenses of doubling the dimension of the basis.

In the construction of the basis we found a new class of spinless objects which generalize the Witten-RSV construction to particles with zero helicity. The new scalar blocks, $m_{n,k}(\alpha, \beta)$, defined for each $k$ sector do not have a direct physical interpretation. The reason is that they each have unphysical poles which come in the form of two irreducible polynomials in the variables $\langle a~b\rangle$ and $[a~b]$ (except when $k=2$ and $k=n-2$ when there is only one such polynomial). We studied the structure of these poles and found that they cancel in pairs with scalar blocks having $k'=k-1$ and $k'=k+1$. This means that the sum over $k$ of the scalar blocks is a physically meaningful object and, in fact, gives the amplitude for a standard bi-adjoint scalar theory. We provided two different formulas for the scalar blocks. Both of them require the solution of polynomial equations. Explicit forms in term of kinematic invariants were provided for $k=2$ and $k=n-2$ but it would be highly desirable to find them for other values of $k$ as well.

The discovery of the BCJ basis was also related to the development of the double-copy connection between Yang-Mills and gravity amplitudes \cite{Bern:2008qj}. The double-copy procedure is, as expected, valid in arbitrary space-time dimensions. It would be very interesting to explore the possibility that the new basis we found in four dimensions could lead to novel relations special to four dimensions.

Finally, our construction gave rise not only to relations among Yang-Mills partial amplitudes but also among a large variety of theories. In fact, any theory that admits a CHY representation can potentially be a part of the novel four-dimensional relations. A particularly interesting example is the application of our general result \eqref{kgen}
\be
A^{\rm target}_{n,k} = \sum_{\alpha\in {{\cal A}_k},\beta\in {{\cal B}_k} }A^{\rm basis}_{n,k}(\alpha)(m_{n,k}^{-1})(\alpha,\beta)A^{\rm basis}_{n,k}(\beta)
\en
to Yang-Mills and gravity. This led to an intrinsically four-dimensional version of the KLT relations, \eqref{kKLT}
\be
M^{\rm gravity}_{n,k}= \sum_{\alpha\in {\cal A}_k,\beta\in {\cal B}_k}A^{\rm YM}_{n,k}(\beta)(m_{n,k}^{-1})(\beta, \alpha)A^{\rm YM}_{n,k}(\alpha).
\en
The KLT relations were originally discovered as a relation among open and closed string amplitudes. It would be interesting to explore the possibility of a stringy derivation of the new four-dimensional version.

\section*{Acknowledgements}

We are grateful to S. He and E. Yuan for useful discussions and comments on the draft. This work is supported by Perimeter Institute for Theoretical Physics. Research at Perimeter Institute is supported by the Government of Canada through Industry Canada and by the Province of Ontario through the Ministry of Research \& Innovation.

\appendix
\section{CHY Formula}

In this appendix we provide a very brief summary of the CHY formulation \cite{CHY}. CHY formulas are a compact way to express complete tree-level amplitudes of scalar, Yang-Mills and gravity theories. In this formulation, every scattering amplitude is expressed by a multidimensional contour integral over the moduli space of $n$ puncture locations $\sigma_a$ on the Riemann sphere:
\be
A=\int d\mu_n\, {\cal I}_L(k,\epsilon,\sigma) {\cal I}_R(k,\epsilon,\sigma)
\en
where the measure
\bes
d\mu_n&=&\frac{d^n \sigma}{{\rm vol}\,{\rm SL}(2,\mathbb{C})}(\sigma_{ij}\sigma_{jk}\sigma_{ki})\prod_{a \neq i,j,k}\delta(\sum_{b \neq a} \frac{k_a\cdot k_b}{\sigma_{ab}})
\tr
&=&\prod_{c\neq p,q,r} d \sigma_c\, (\sigma_{pq}\sigma_{qr}\sigma_{rp})(\sigma_{ij}\sigma_{jk}\sigma_{ki})\prod_{a \neq i,j,k}\delta(\sum_{b \neq a} \frac{k_a\cdot k_b}{\sigma_{ab}})
\ens
and the integrand ${\cal I}(k,\epsilon,\sigma)$ depends on polarization vectors $\epsilon^\mu_a$, momenta $k^\mu_a$ and puncture locations, $\sigma_a$.

Integrands for the three theories we used in this work were presented in section 2 but we repeat them here for the reader's convenience.

For the standard bi-adjoint scalar $\phi^3$ theory the integrand is given by \eqref{scalarI},
\be
{\cal I}_L^{\phi^3}(\alpha|\beta)=C_n(\alpha), \quad {\cal I}_R^{\phi^3}(\alpha|\beta) = C_n(\beta)
\en
with $C_n(\alpha)=1/(\sigma_{\alpha_1,\alpha_2}\sigma_{\alpha_2,\alpha_3}\cdots \sigma_{\alpha_n,\alpha_1})$.

For Yang-Mills theories, the integrand is given by \eqref{YMI},
\be
{\cal I}^{\rm YM}_L(\alpha)=C_n(\alpha),\quad {\cal I}^{\rm YM}_R = {\rm Pf'}\Psi_n(\epsilon,k,\sigma)
\en
while for gravity amplitudes one has \eqref{gravityI},
\be
{\cal I}^{\rm YM}_L(\alpha)={\rm Pf'}\Psi_n(\epsilon,k,\sigma),\quad {\cal I}^{\rm YM}_R = {\rm Pf'}\Psi_n(\epsilon,k,\sigma).
\en
In both Yang-Mills and gravity amplitudes ${\rm Pf'}\Psi_n$ denotes the reduced Pfaffian of a $2n\times 2n$ matrix $\Psi$. The precise definition is the following. ${\rm Pf'}\Psi_n=-\frac{(-1)^{a+b}}{\sigma_{ab}}{\rm Pf} [\Psi_n]_{\hat{a},\hat{b}}$. This turns out to be a permutation invariant quantity constructed using the submatrix of $\Psi_n$ obtained by deleting columns and rows denoted by $\{a,b\}$. And $\Psi_n$ is defined as
\begin{equation}\label{Psi}
{\setstretch{1.4}
\Psi_n = \left(
         \begin{array}{c|c}
           ~~A_n~~ &  -C_n^{\rm T} \\
           \hline
           C_n & B_n \\
         \end{array}
       \right)
}
\end{equation}
where $A$, $B$ and $C$ are $n\times n$ matrices:
\be
A_{ab} = \begin{cases} \displaystyle \frac{k_{a}\cdot k_b}{\sigma_{ab}} & a\neq b,\\
\displaystyle  0 & a=b,\end{cases} \quad B_{ab} = \begin{cases} \displaystyle \frac{\epsilon_a\cdot\epsilon_b}{\sigma_{ab}} & a\neq b,\\
\displaystyle  0 & a=b,\end{cases}\quad
C_{ab} = \begin{cases} \displaystyle \frac{\epsilon_a \cdot k_b}{\sigma_{ab}} &\quad a\neq b,\\
\displaystyle -\hspace{-0.5em}\sum_{c=1,\,c\neq a}^n \hspace{-0.5em}C_{ac} &\quad a=b.\end{cases}
\label{ABCmatrix}
\en


%

\section{Bi-adjoint Scalar}

Here we give a short review of how to compute bi-adjoint scalar amplitudes in terms of Mandelstam invariants, which is also explained in \cite{Cachazo:2013iea}. The definition for it is:
\be
m_n(\alpha,\beta)=\sum_{{\sigma}\in {\rm solutions}}\frac{{C}_n(\alpha){C}_n(\beta)}{{\rm det'}\Phi}.
\en
Some simple examples for $n=5$ are
\bes
m_5(12345|12345)&=&\frac{1}{s_{12}s_{34}}+\frac{1}{s_{23}s_{45}}+\frac{1}{s_{34}s_{51}}+\frac{1}{s_{45}s_{12}}+\frac{1}{s_{54}s_{23}},\\
m_5(12345|13245)&=&-\frac{1}{s_{23}s_{45}}-\frac{1}{s_{23}s_{51}},\quad m_5(12345|13524)=0.
\ens

For $n=6$ we include the cases that are useful in the computation of the $4\times 4$ matrix $m_6$ used in section 3:
\bes
&\,&m_6(124356|153462)=\frac{1}{s_{12}s_{126}s_{34}}+\frac{1}{s_{12}s_{126}s_{35}},\\
&\,&m_6(124356|154362)=-\frac{1}{s_{12}s_{34}s_{126}},\quad m_6(124356|152463)=0.
\ens
From these examples we can see that $m_6(\alpha|\beta)$ is a sum over double flavor-ordered trivalent graphs, each of which is evaluated by the product of its propagators. As a result, we have the following:
\be
m_n(\alpha|\beta)=\pm \sum_{g\in T(\alpha)\bigcap T(\beta)}\prod_{e\in E(g)}\frac{1}{s_e}
\en
with $T(\alpha)$ denoted as the set of $\alpha$-color-ordered diagrams and $E(g)$ the set of inner edges of each diagram. $s_e=P_e^2$ where $P_e$ is the momentum flowing along the edge $e$ in the set of edges $E(g)$. The overall sign can be fixed numerically which is also defined in \cite{Cachazo:2013iea}. In particular, if $T(\alpha)\bigcap T(\beta)=\varnothing$ we have $m_n(\alpha|\beta)=0$.

\end{document}